\documentclass{PoS}

\usepackage{subfigure}
\usepackage[UKenglish]{babel}
\usepackage{natbib}
\pdfoutput=1

\newcommand{\gev}{\, {\rm GeV}}
\newcommand{\mev}{\, {\rm MeV}}
\newcommand{\be}{\begin{equation}}
\newcommand{\ee}{\end{equation}}
\newcommand{\bea}{\begin{eqnarray}}
\newcommand{\eea}{\end{eqnarray}}

\title{A $N_f = 2 + 1 + 1$ ``twisted'' determination of \\the $b$-quark mass, $f_{B}$ and $f_{B_s}$}

\ShortTitle{A $N_f = 2 + 1 + 1$ ``twisted'' determination of the $b$-quark mass, $f_{B}$ and $f_{B_s}$}

\author{
N. Carrasco$^{(a,b)}$,  P. Dimopoulos$^{(c,d)}$,  R. Frezzotti$^{(d)}$,  V. Gim\'enez$^{(a)}$,  P. Lami$^{(e,b)}$, 

V. Lubicz$^{(e,b)}$, \speaker{E. Picca}$^{(e,b)}$, L. Riggio$^{(e,b)}$, G.C. Rossi$^{(d)}$, F. Sanfilippo$^{(f)}$, S. Simula$^{(b)}$, C. Tarantino$^{(e,b)}$\\
\\
\llap{$^{(a)}$}Departament de F\'isica Te\`orica and IFIC, Univ. de Val\`encia-CSIC, Av. Dr. Moliner 50, E-46100 Val\`encia, Spain, E-mail: \email{nuria.carrasco@uv.es}, \email{vincente.gimenez@uv.es}\\
\llap{$^{(b)}$}INFN, Sezione di Roma Tre, Via della Vasca Navale 84, I-00146 Rome, Italy, E-mail: \email{simula@roma3.infn.it}\\
\llap{$^{(c)}$}Centro Fermi - Museo storico della Fisica e Centro Studi e Ricerche Enrico Fermi, Compendio del Viminale, Piazza del Viminale 1, I-00184 Rome, Italy\\
\llap{$^{(d)}$}Dipartimento di Fisica, Universit\`a di Roma ``Tor Vergata'' and INFN Sezione ``Tor Vergata'', Via della Ricerca Scientifica 1, I-00133 Rome, Italy, E-mail: \email{dimopoulos@roma2.infn.it}, \email{frezzotti@roma2.infn.it}, \email{rossig@roma2.infn.it}\\
\llap{$^{(e)}$}Dipartimento di Matematica e Fisica, Universit\`a Roma Tre, Via della Vasca Navale 84, I-00146 Rome, Italy, Email: \email{lamipaolo@gmail.com}, \email{lubicz@fis.uniroma3.it}, \email{e.picca88@gmail.com}, \email{lorenzo.riggio@gmail.com}, \email{tarantino@fis.uniroma3.it}\\
\llap{$^{(f)}$}Laboratoire de Physique Th\'eorique (B$\hat{a}$t.~210), Universit\'e Paris Sud, F-91405 Orsay-Cedex, France, Email: \email{francesco.sanfilippo@th.u-psud.fr}
\vskip+0.5cm \textbf{For the ETM Collaboration}}

\abstract{We present a lattice QCD determination of the $b$-quark mass and of the $f_{B_s}$ and $f_B$ decay constants performed with $N_f = 2 + 1 + 1$ twisted mass Wilson fermions. We have used simulations at three values of the lattice spacing generated by ETMC with pion masses ranging from 210 to 440 MeV. To obtain physical quantities we performed a combined chiral and continuum limit and an extrapolation in the heavy quark mass from the charm to the $b$-quark region using suitable ratios calculated at nearby heavy-quark masses having an exactly known static limit. Our results are: $m_b(m_b) = 4.29 (13)$ GeV, $f_B = 196 (9)$ MeV, $f_{B_s} = 235 (9)$ MeV, $f_{B_s} / f_B = 1.201 (25)$, $(f_{B_s}/f_B)/(f_K/f_\pi) = 1.007 (16)$ and
$(f_{B_s}/f_B)/(f_{D_s}/f_D) = 1.008 (13)$.}

\FullConference{31st International Symposium on Lattice Field Theory LATTICE 2013\\
                 July 29 - August 3, 2013\\
                 Mainz, Germany}

\begin{document}

\section{Introduction}
The study of physical processes involving the $b$ quark are of utmost importance for accurate tests of the Standard Model (SM) and for searching New Physics (NP) effects.

In spite of the great success of the SM in describing the observed processes, the SM presents the unsatisfactory aspect of having several parameters with unpredicted values. In the quark sector only, these are the six quark masses and the four parameters of the Cabibbo-Kobayashi-Maskawa (CKM) matrix. Quark masses are naturally expected to be of the order of the Higgs boson's vacuum expectation value (VEV), but, except for the top mass, they are found to be several orders of magnitude smaller than expected. Furthermore quark masses show a hierarchy whose origin is not understood. In order to unravel these aspects we need accurate values of the quark masses that can be calculated using Lattice QCD. 

Other physical quantities of interest are the decay constants that allow us to extract the CKM elements using experimental results, and also to probe NP. In particular, in B-physics, the theoretical predictions for the two NP-sensitive processes $B\rightarrow\tau\nu$ and $B^0_s\rightarrow\mu^+\mu^-$ require a precise knowledge of the pseudoscalar (PS) decay constants $f_B$ and $f_{B_s}$, respectively.

In this contribution we report the results for the $b$ quark mass $m_b$ and for the $f_B$ and $f_{B_s}$ decay constants (as well as for their ratio $f_{B_s}/f_B$) obtained adopting Wilson twisted mass fermions at maximal twist \cite{Frezzotti:2003xj} and using the gauge configurations produced by European Twisted Mass Collaboration (ETMC) with $N_f=2+1+1$ dynamical sea quarks \cite{Baron:2010bv}. In order to deal with heavy mesons on the lattice we used smeared interpolating  operators and we performed an extrapolation in the heavy quark mass from the charm to the $b$-quark region, using the {\em ratio method} of ref.~\cite{Blossier:2009hg}. Further details of the analysis procedure can be found in refs.~\cite{Carrasco:2013zta,Carrasco:2012de,Dimopoulos:2011gx}.

\section{Simulation details and lattice set up}
We have used the $N_f=2+1+1$ gauge configurations generated by ETMC \cite{Baron:2010bv} at three values of the lattice spacing corresponding to $\beta=\{1.90,~1.95,~2.10\}$. Subsets of well-separated trajectories were selected to avoid autocorrelations. Bare quark masses are chosen so that the pion mass, extrapolated to the continuum and infinite volume limits, range from $\simeq 210$ to $\simeq 440$ MeV. In \cite{proc_1} we have determined the lattice spacing $a$ and the physical light $m_{u/d}$ and strange $m_s$ quark masses. Since those results were not available at the time of the present analysis, we have used the following preliminary results: $a=\{0.0886(27),0.0815(21),0.0619(11)\}$~fm, $m_{u/d}=3.7(1)$ GeV and $m_s=100(3)$~MeV in the $\overline{MS}$ scheme at 2 GeV.

In the calculation of PS heavy-light meson masses we used smeared and optimized interpolating fields as described in~\cite{Carrasco:2013zta} in order to keep the noise-to-signal ratio under control. The optimal fields are a linear superposition of smeared and local fields and are determined by tuning the coefficients of the combination numerically. The smearing technique and the optimal fields reduce the overlap of the interpolating field with the excited states and allow to extract meson masses at relatively small Euclidean time. For the decay constants we used only the smearing technique because all the sink/source combinations of smeared and local fields necessary for the optimal operators were not available. The results for $f_B$ and $f_{B_s}$ will be updated once these data will be produced.

Renormalized quark masses $\bar{\mu}$ are obtained from the bare ones $\mu$ using the renormalization constant (RC) $Z_{\mu}=Z_{P}^{-1}$ calculated in the RI-MOM scheme with two methods, labelled as M1 and M2 (see ref.~\cite{ETM:2011aa}), which differ by ${\cal{O}}(a^2)$ effects. The difference in the results obtained in the continuum limit is used to estimate the systematic uncertainty related to the RC $Z_P$.

\section{The $b$-quark mass}
\begin{figure}[htb!]
\centering
\subfigure[]{\includegraphics[scale=0.26]{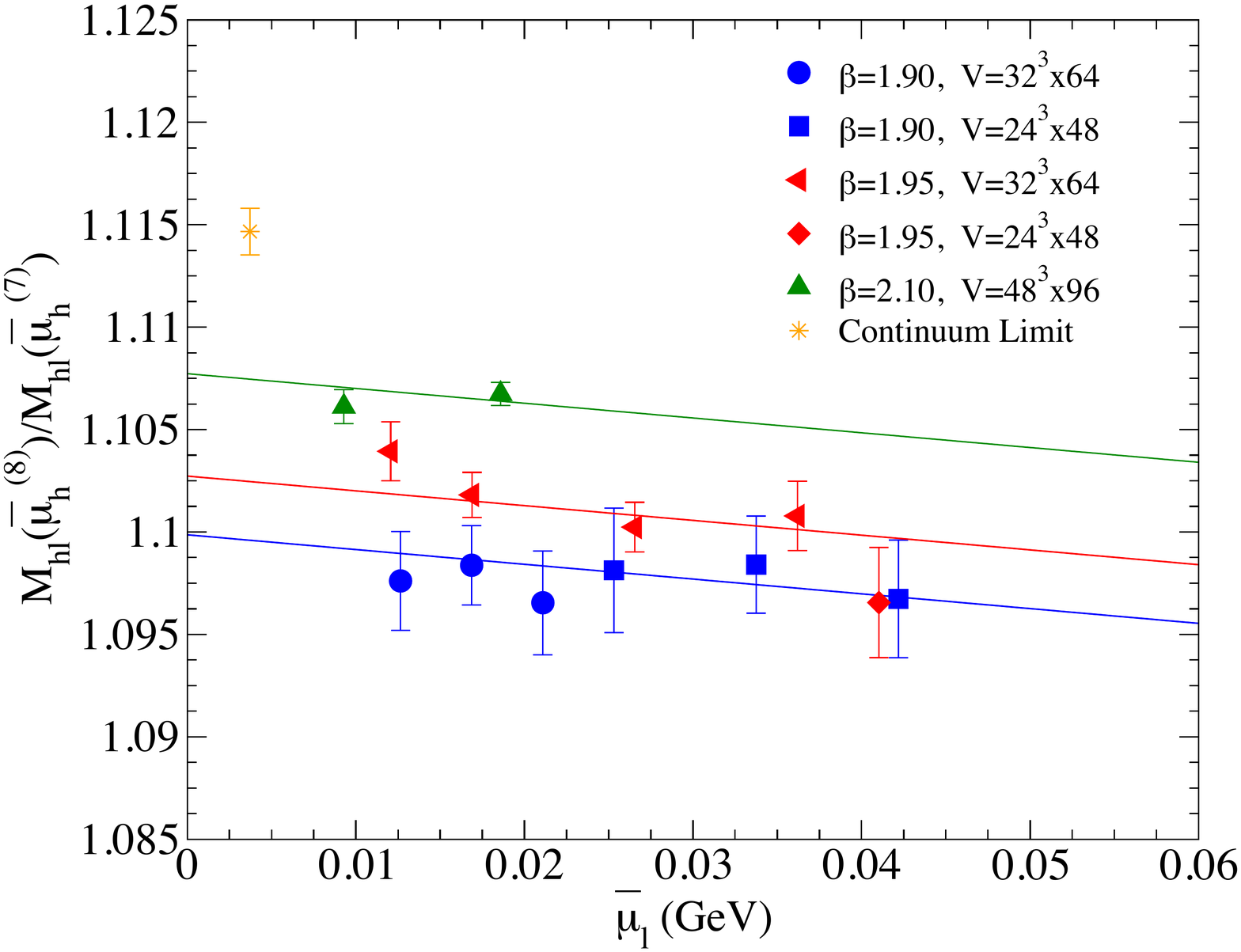}\label{fig_fitmb:y}}
\subfigure[]{\includegraphics[scale=0.26]{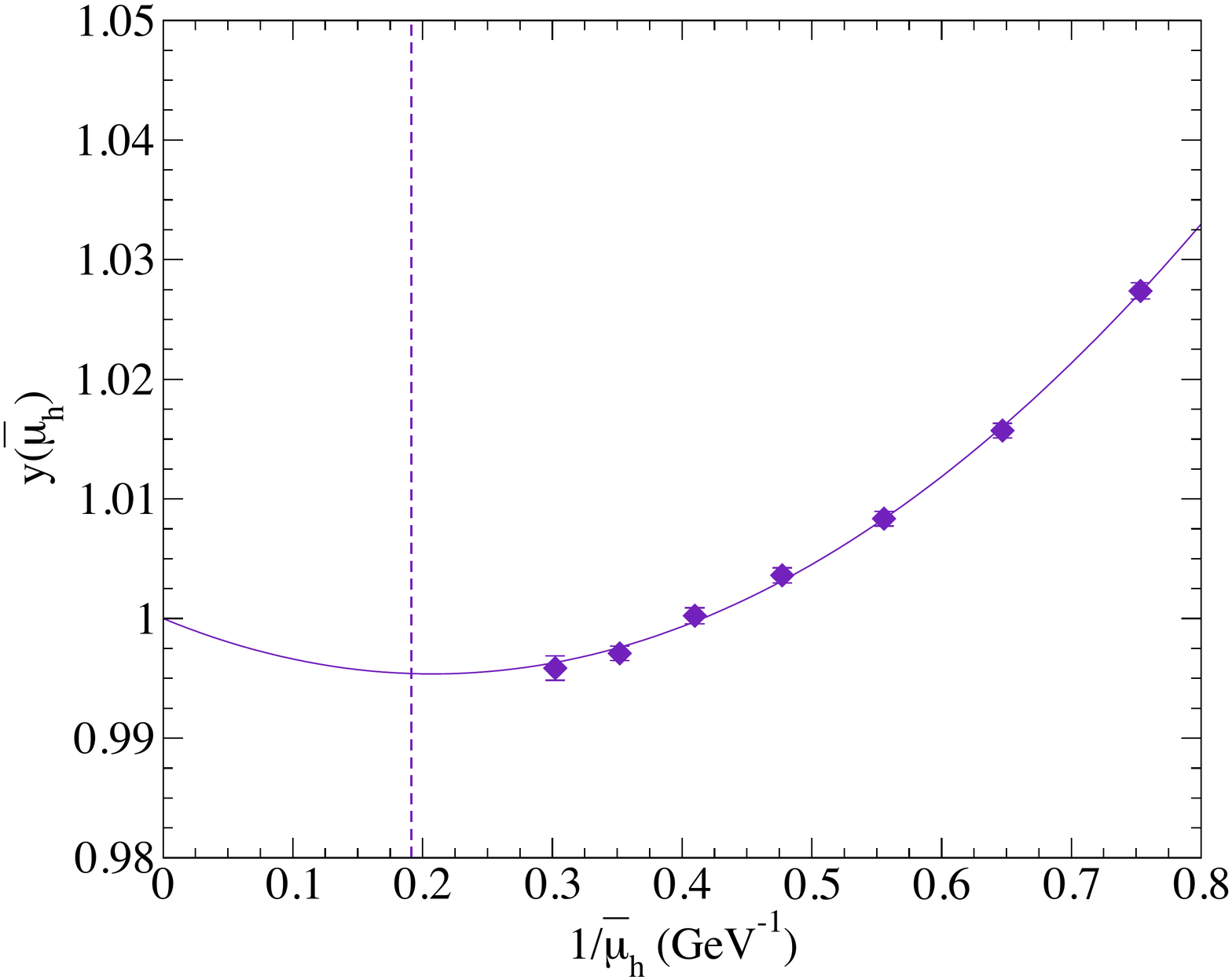}\label{fig_fitmb_mh}}
\caption{{\it (a) Chiral and continuum fit of the ratio of the heavy-light PS meson masses for our largest values of the heavy-quark mass versus the light-quark mass $\bar{\mu}_\ell$. (b) $y(\bar{\mu}_h)$ against $1/\bar{\mu}_h$ using eq.~(\ref{eq_ansatz}). The vertical line represents the location of the inverse $b$-quark mass $1 / \bar{\mu}_b$.}}
\label{fig_fitmb}
\end{figure}

The $b$-quark mass is calculated using the ratio method of ref.~\cite{Blossier:2009hg}. The first step consists in considering an appropriate sequence of heavy-quark masses $(\bar{\mu}_h^{(1)},\bar{\mu}_h^{(2)}, ...,\bar{\mu}_h^{(N)})$ with fixed ratio $\lambda$ between any two successive values $\bar{\mu}_h^{(n)}=\lambda\bar{\mu}_h^{(n-1)}$. The ratio $\lambda$ is chosen in such a way that after a finite number of steps the heavy-light PS meson mass assumes the experimental value $M_B = 5.279$ GeV. The Heavy Quark Effective Theory (HQET) suggests the following asymptotic behaviour
\begin{equation}\label{eq_massahqet}
\lim_{\mu_h^{pole}\rightarrow\infty}\left(\frac{M_{h\ell}}{\mu_h^{pole}}\right)=\mbox{constant},
\end{equation}
where $M_{h\ell}$ is the heavy-light PS meson mass and $\mu_h^{pole}$ is the heavy-quark pole mass. The latter is related to the (renormalized) heavy-quark mass $\bar{\mu}_h$, given in the $\overline{MS}$ scheme, by $\mu_h^{pole}=\rho(\bar{\mu}_h,\nu)\bar{\mu}_h(\nu)$ with $\nu$ being the renormalization scale. The function $\rho(\bar{\mu}_h,\nu)$ is known up to $N^3LO$ in perturbation theory (PT). Thus we defined the ratios $y(\bar{\mu}_h,\lambda;\bar{\mu}_\ell,a^2)$ as
\begin{equation}\label{eq_y}
y(\bar{\mu}_h,\lambda;\bar{\mu}_\ell,a^2) = \frac{M_{h\ell}(\bar{\mu}_h^{(n)})}{\mu_h^{(n) pole}} \frac{\mu_h^{(n-1) pole}}{M_{h\ell}(\bar{\mu}_h^{(n-1)})} = \lambda^{-1}\frac{M_{h\ell}(\bar{\mu}_h^{(n)})}{M_{h\ell}(\bar{\mu}_h^{(n)}/\lambda)} 
                                                                      \frac{\rho(\bar{\mu}^{(n)}_h / \lambda,\nu)}{\rho(\bar{\mu}^{(n)}_h,\nu)} ~ ,
\end{equation} 
where $\bar{\mu}_\ell$ is the light quark mass. We calculated the chiral and continuum extrapolation $y(\bar{\mu}_h) \equiv y(\bar{\mu}_h, \lambda;$ $\bar{\mu}_{u/d}, a^2 = 0)$ by performing a simultaneous fit in the light quark mass and in the lattice spacing for each value of $\bar{\mu}_h$. We considered a linear dependence on $\bar{\mu}_\ell$ and $a^2$, accounting for the automatic ${\cal{O}}(a)$ improvement of the twisted mass action at maximal twist. The result of the fit at the largest value of the heavy-quark mass is reported in fig.~\ref{fig_fitmb:y}.

Using eq.~(\ref{eq_massahqet}) the ratio $y(\bar{\mu}_h)$ has an exactly known static limit, namely: $\lim_{\bar{\mu}_{h}\rightarrow\infty}y(\bar{\mu}_h)=1$, so that the dependence of the ratio $y(\bar{\mu}_h)$ on the heavy-quark mass $\bar{\mu}_h$ can be written as
\begin{equation}\label{eq_ansatz}
y(\bar{\mu}_h)=1+\frac{\eta_1}{\bar{\mu}_h}+\frac{\eta_2}{\bar{\mu}_h^2} ~ .
\end{equation}
The corresponding fit of the lattice data is presented in fig.~\ref{fig_fitmb_mh}. Finally the value for the $b$-quark mass is computed from the {\em chain} equation
\begin{equation}\label{eq_chain}
y(\bar{\mu}_h^{(2)})y(\bar{\mu}_h^{(3)})...y(\bar{\mu}_h^{(K+1)})=\lambda^{-K}\frac{M_{h,u/d}(\bar{\mu}_h^{(K+1)})}{M_{h,u/d}(\bar{\mu}_h^{(1)})}\left[\frac{\rho(\bar{\mu}_h^{(1)},\nu)}{\rho(\bar{\mu}_h^{(K+1)},\nu)}\right],
\end{equation}
where the values for $\lambda$, $K$ and $\bar{\mu}_h^{(1)}$ are chosen in such a way that $M_{h,u/d}(\bar{\mu}_h^{(K+1)})$ is equal to the experimental value of the B meson mass.  In eq.~(\ref{eq_chain}) the quantity $M_{h,u/d}(\bar{\mu}_h^{(1)})$ is the result of the combined chiral and continuum fit of the PS meson mass evaluated at the reference heavy quark mass $\bar{\mu}_h^{(1)}$. We found $(\bar{\mu}_h^{(1)},\lambda,K)=(1.14~\gev,1.1644,10)$ from which the $b$-quark mass $\bar{\mu}_b=\lambda^{K}\bar{\mu}^{(1)}$, in the $\overline{MS}$ scheme at $\nu = 2$ GeV, resulted to be $\bar{\mu}_b=5.22(10) \gev$, where the error comes from the statistical uncertainties of the lattice data, the fitting error and the statistical uncertainties on the lattice spacing and the renormalization constant of the method M1. Using the set of RCs from the method M2 we obtained a result for $\bar{\mu}_b$ lower by $\simeq 3\%$. We took the average of the two results and we considered half of the difference as the systematic uncertainty due to the RCs. 
Evolving from $2 \gev$ to $m_b$ using $\Lambda_{QCD}^{N_f=4} = 296 \mev$ (see \cite{Carrasco:2013zta} and references therein), we obtain
\begin{equation} \label{eq_mb}
m_b(m_b) = 4.29(8)(5)(9) \gev = 4.29(13) \gev ~ ,
\end{equation}
where the first error is the statistical+fitting error, the second and the third ones are the systematic uncertainties on the renormalization constants and lattice spacing, respectively. The estimate of the latter uncertainty is still preliminary. The total error is the sum in quadrature of the uncertainties. Our result (\ref{eq_mb}) is well compatible with the one found with $N_f = 2$~\cite{Carrasco:2013zta,Carrasco:2013proc}, indicating that the effect of the strange and charm sea quarks is not visible at the present level of accuracy.

\section{Decay constants}
\begin{figure}[htb!]
\centering
\subfigure[]{\includegraphics[scale=0.26]{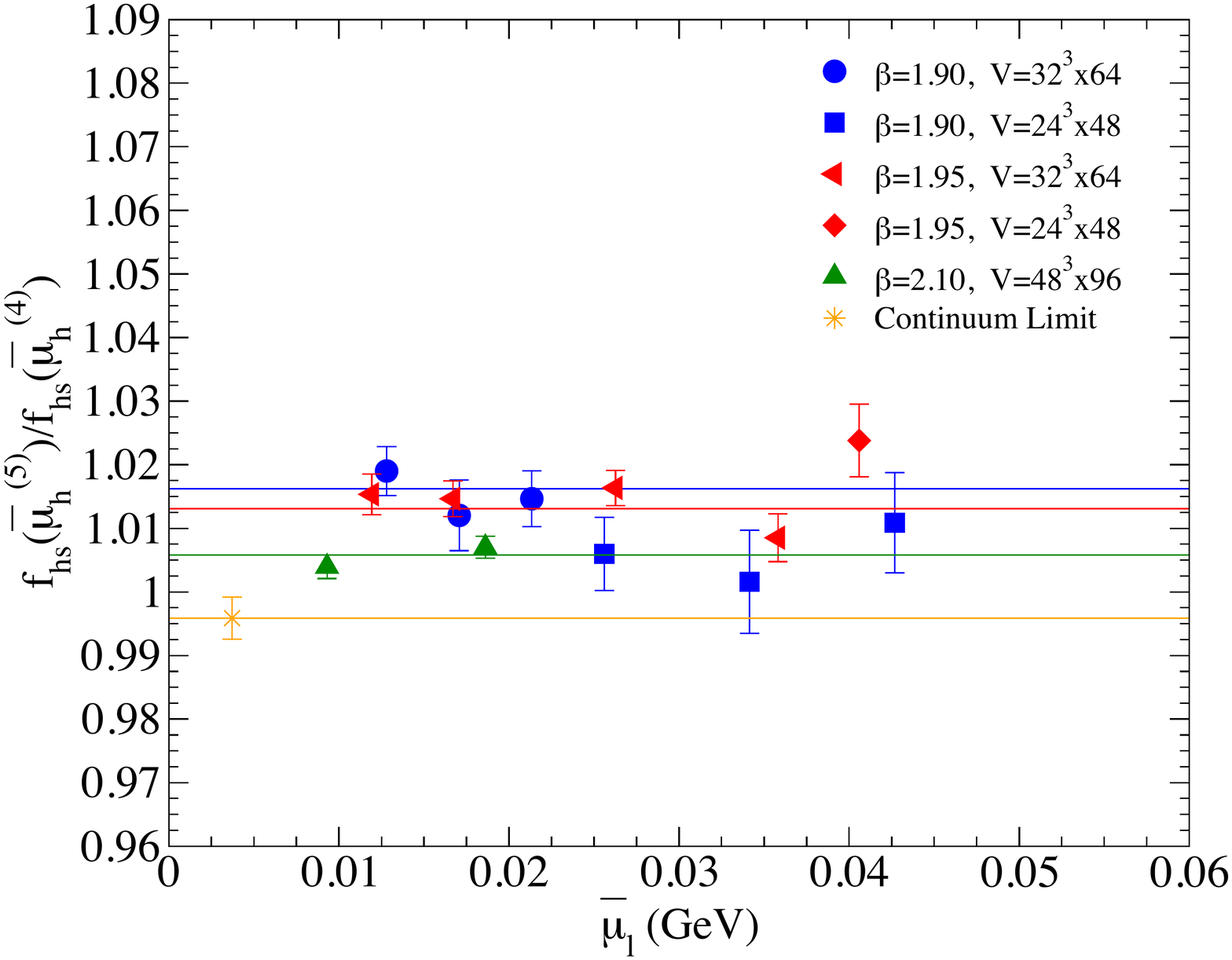}\label{fig_frapp}}
\subfigure[]{\includegraphics[scale=0.26]{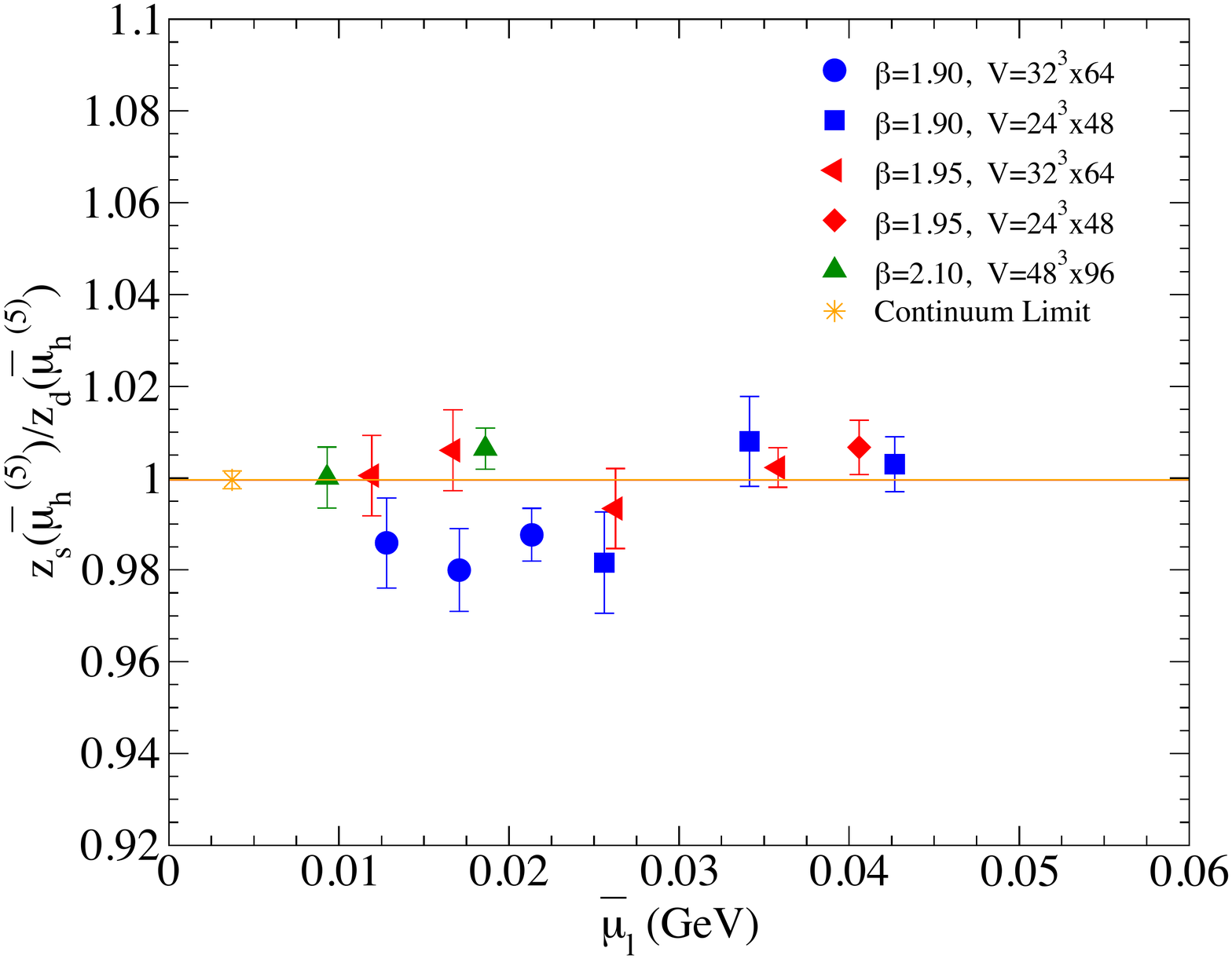}\label{fig_fidrapp}}
\caption{{\it Chiral and continuum fit of the ratio of the decay constant (a) and of the double ratio (b) for our largest values of the heavy-quark mass as a function of $\bar{\mu}_\ell$. A linear fit both in $\bar{\mu}_\ell$ and $a^2$ has been adopted.}}\label{fig_fitf1}
\end{figure}

We used a similar strategy for the calculation of the decay constant $f_{B_s}$ and the ratio $f_{B_s}/f_B$, from which we obtained $f_B$. We considered the HQET predictions: $\lim_{\mu_h^{pole}\rightarrow\infty}f_{hs} \sqrt{\mu_h^{pole}} = \mbox{constant}$ and $\lim_{\mu_h^{pole}\rightarrow\infty}(f_{hs}/f_{h\ell}) = \mbox{constant}$,
where $f_{hs} (f_{h\ell})$ is the heavy-strange(light) PS meson decay constant. We defined the ratios 
\begin{eqnarray}\label{eq_frappdefd}
z_d(\bar{\mu}_h,\lambda;\bar{\mu}_\ell,a^2) & = & \sqrt{\lambda}\frac{f_{h\ell}(\bar{\mu}_h,\bar{\mu}_\ell,a^2)}{f_{h\ell}(\bar{\mu}_h/\lambda,\bar{\mu}_\ell,a^2)}\cdot\frac{C_A^{stat}(\nu^{*},\bar{\mu}_h/\lambda)}{C_A^{stat}(\nu^{*},\bar{\mu}_h)}\frac{\rho(\bar{\mu}_h,\nu)^{1/2}}{\rho(\bar{\mu}_h/\lambda,\nu)^{1/2}} ~ , \\[2mm] \label{eq_frappdefs}
z_s(\bar{\mu}_h,\lambda;\bar{\mu}_\ell,\bar{\mu}_s,a^2) & = & \sqrt{\lambda}\frac{f_{hs}(\bar{\mu}_h,\bar{\mu}_\ell,\bar{\mu}_s,a^2)}{f_{hs}(\bar{\mu}_h/\lambda,\bar{\mu}_\ell,\bar{\mu}_s,a^2)}\cdot\frac{C_A^{stat}(\nu^{*},\bar{\mu}_h/\lambda)}{C_A^{stat}(\nu^{*},\bar{\mu}_h)}\frac{\rho(\bar{\mu}_h,\nu)^{1/2}}{\rho(\bar{\nu}_h/\lambda,\nu)^{1/2}},
\end{eqnarray}
where the factor $C_A^{stat}(\nu^{*},\bar{\mu}_h)$, known up to $N^2LO$ in PT, provides the matching between the decay constant in QCD and its static-light counterpart in HQET (the arbitrary renormalization scale $\nu^{*}$ of HQET cancels out in the ratio). From eqs.~(\ref{eq_frappdefd}-\ref{eq_frappdefs}) we form the double ratio
\begin{equation}\label{eq_drapp}
\zeta(\bar{\mu}_h,\lambda;\bar{\mu}_\ell,\bar{\mu}_s,a^2)=\frac{z_s(\bar{\mu}_h,\lambda;\bar{\mu}_\ell,\bar{\mu}_s,a^2)}{z_d(\bar{\mu}_h,\lambda;\bar{\mu}_\ell,a^2)}.
\end{equation}
By construction the ratios $z_d$, $z_s$ and $\zeta$ have an exactly known static limit equal to unity. They exhibit a smooth chiral and continuum behavior, as shown in fig.~\ref{fig_fitf1} at our largest heavy-quark mass. As mentioned before, in this case we couldn't use the optimized interpolating operators, so that we considered data up to the mass $\bar{\mu}_h^{(5)}$, because the signals for both the PS meson masses and the decay constants corresponding to heavier values of $\bar{\mu}_h$ were too noisy.

\begin{figure}[htb!]
\centering
\subfigure[]{\includegraphics[scale=0.26]{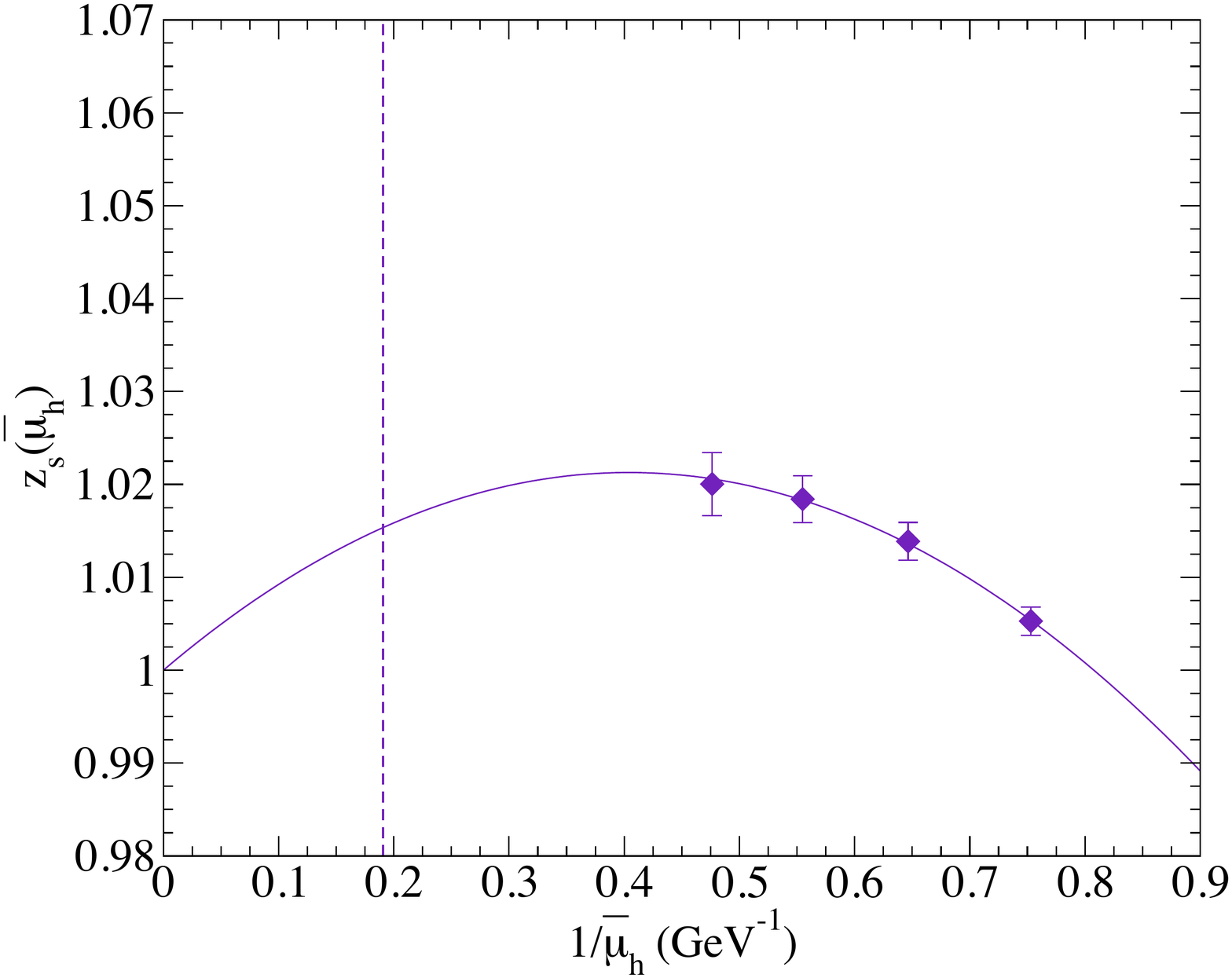}\label{fig_mh_frapp}}
\subfigure[]{\includegraphics[scale=0.26]{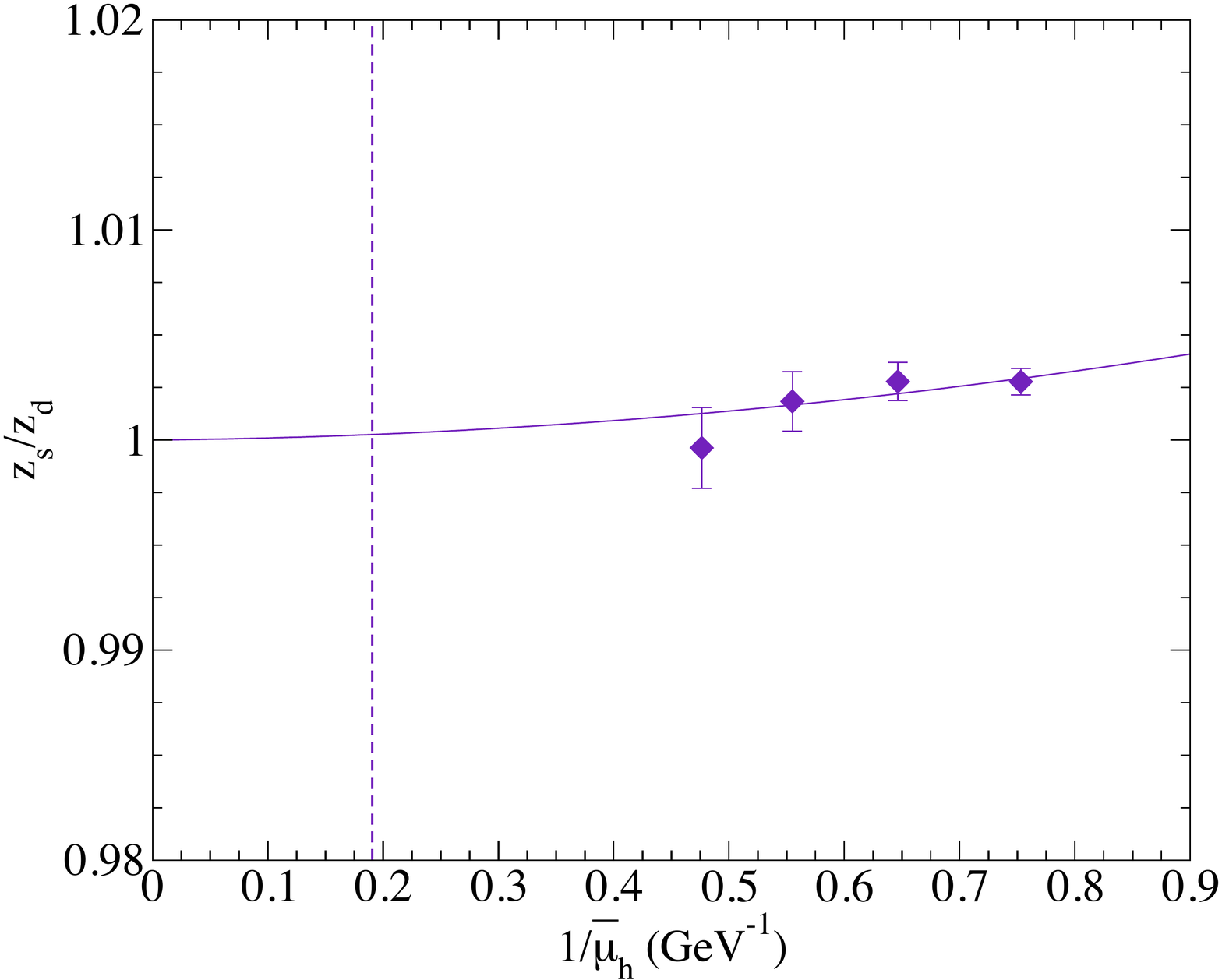}\label{fig_mh_fidrapp}}
\caption{{\it Heavy quark mass dependence of the ratio $z_s$ and of the double ratio $z_s/z_d$. The vertical lines represent the location of the inverse $b$-quark mass $1 / \bar{\mu}_b$.}}
\label{fig_fitf2}
\end{figure}

The results for $f_{B_s}$ and $f_{B_s}/f_B$ are determined from the chain equations
\begin{eqnarray}
z_s(\bar{\mu}_h^{(2)})z_s(\bar{\mu}_h^{(3)})~...~z_s(\bar{\mu}_h^{(K+1)}) & = & \lambda^{K/2}~\frac{f_{hs}(\bar{\mu}_h^{(K+1)})}{f_{hs}(\bar{\mu}_h^{(1)})}\cdot\frac{C_A^{stat}(\nu^{*},\bar{\mu}_h^{(1)})}{C_A^{stat}(\nu^{*},\bar{\mu}_h^{(K+1)})}\sqrt{\frac{\rho(\bar{\mu}_h^{(K+1)},\nu)}{\rho(\bar{\mu}_h^{(1)},\nu)}}, \\[2mm]
\zeta(\bar{\mu}_h^{(2)})\zeta(\bar{\mu}_h^{(3)})~...~\zeta(\bar{\mu}_h^{(K+1)}) & = & \lambda^{K/2} \frac{f_{hs}(\bar{\mu}_h^{(K+1)})/f_{hu/d}(\bar{\mu}_h^{(K+1)})}{f_{hs}(\bar{\mu}_h^{(1)})/f_{hu/d}(\bar{\mu}_h^{(1)})},
\end{eqnarray}
where $\bar{\mu}_h^{(1)}$, $\lambda$ and $K$ are the values found in the $b$-quark mass analysis and $z_s(\bar{\mu}_h^{(n)})$ and $\zeta(\bar{\mu}_h^{(n)})$ are calculated after the chiral and continuum extrapolations. For the decay constant ratio $f_{hs}/f_{h\ell}$ heavy meson chiral perturbation theory (HMChPT) predicts at NLO a linear+logarithmic dependence on the light quark mass. We took advantage of the fact that the double ratio of the decay constants
\begin{equation} \label{eq_Rf}
R_f=[(f_{hs}/f_{h\ell})/(f_{s\ell}/f_{\ell \ell})] 
\end{equation}
exhibits a large cancellation of the chiral logs. Thus we extrapolated the quantity (\ref{eq_Rf}) to the chiral and continuum limits, and by using the result $f_K/f_{\pi}=1.193(16)$ from \cite{proc_2} we extracted the ratio $f_{B_s}/f_B$. For the chiral extrapolation we used either a linear fit or the one suggested by combining SU(2) ChPT and HMChPT, namely:
\begin{equation} \label{eq_chiral}
\begin{array}{ll}
\mbox{Lin.:      }~~~~~~~~~~~~~~~~~R_f = a_h^{(1)}+b_h^{(1)}\bar{\mu}_\ell+D_h^{(1)}a^2 ~ , \\[2mm]
\mbox{HMChPT:      }~~~~~~~~R_f = a_h^{(2)}\left[1+b_h^{(2)}\bar{\mu}_\ell+\left[\frac{3(1+3\hat{g}^2)}{4}-\frac{5}{4}\right]\frac{2B_0\bar{\mu}_\ell}{(4\pi f_0)^2}\log{\left(\frac{2B_0\bar{\mu}_\ell}{(4\pi f_0)^2}\right)}\right] +D_h^{(2)}a^2 ~ ,
\end{array}
\end{equation} 
where $\hat{g}=0.61(7)$ (see \cite{Carrasco:2013zta} and references therein). 
\begin{figure}[htb!]
\centering
\includegraphics[scale=0.28]{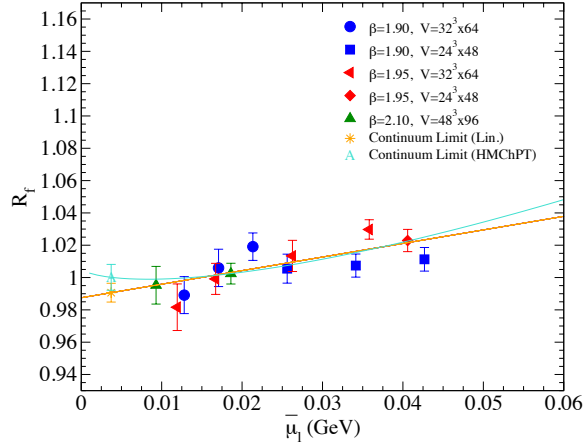}
\caption{{\it Fit for the ratio $R_f$ at the reference heavy quark mass $\bar{\mu}_h^{(1)}$ using $f_K/f_{\pi}$ . The solid lines correspond to linear and HMChPT fits (see Eq.~(\ref{eq_chiral})).}}
\label{fig_fitftrig}
\end{figure}
The results of the two fits are presented in fig.~\ref{fig_fitftrig}, where it can clearly be seen that the two values of $R_f$ at the physical pion point are compatible within the errors. We took their average as our final result and we considered half of their difference as a systematic uncertainty. Finally we performed a fit of the ratio (\ref{eq_frappdefs}) and of the the double ratio (\ref{eq_drapp}) using the same functional form as in eq.~(\ref{eq_ansatz}) in the inverse heavy-quark mass. The results are illustrated in fig.~\ref{fig_fitf2}.
At the physical point in the isospin symmetric limit we obtain
\begin{eqnarray} \label{eq_resultsBsBKPi}
(f_{B_s} / f_B) / (f_K / f_\pi) & = & 1.007(15)(5) = 1.007(16) ~ , \\ \label{eq_resultsBsBDsD}
(f_{B_s} / f_B) / (f_{D_s} / f_D) & = & 1.008(13)(1) = 1.008(13) ~ , \\ \label{eq_resultsBsB}
f_{B_s}/f_B & = & 1.201(24)(6) =1.201(25) ~ , \\ \label{eq_resultsBs}
f_{B_s} & = & 235(8)(5) \mev = 235(9) \mev ~ , \\ \label{eq_resultsB}
f_B & = &196(8)(4) \mev =196(9) \mev ~ , 
\end{eqnarray}
where the first error comes from the statistical uncertainties and fitting error and the statistical uncertainties on the lattice spacing and the mass renormalization constants (for $f_{B_s}/f_B$ and $f_B$ the uncertainty on $f_K/f_{\pi}$ is also included), while the second error comes from the chiral extrapolations (\ref{eq_chiral}) for $(f_{B_s} / f_B) / (f_K / f_\pi)$ and $f_{B_s}/f_B$, from the uncertainty on the $b$-quark mass for $(f_{B_s} / f_B) / (f_{D_s} / f_D)$ and from the systematic uncertainty on the lattice spacing for the decay constants. The systematic uncertainty due to the RCs turns out to be negligible. 

Our results (\ref{eq_resultsBsB}-\ref{eq_resultsB}) obtained with $N_f = 2+1+1$ are consistent with the FLAG averages \cite{FLAG2} corresponding to both $N_f = 2$ and $N_f = 2+1$.
Notice the remarkable results for the double ratios (\ref{eq_resultsBsBKPi}) and (\ref{eq_resultsBsBDsD}), which imply that SU(3) breaking effects in the ratio of PS meson decay constants are the same in the light, charm and bottom sectors within a percent accuracy.

\section*{Acknowledgements}
We acknowledge the CPU time provided by the PRACE Research Infrastructure under the project PRA067 at the J\"ulich  and CINECA SuperComputing Centers, and by the agreement between INFN and CINECA under the specific initiative INFN-RM123.


\end{document}